\documentclass[aps,prb,twocolumn,showpacs,superscriptaddress,groupedaddress]{revtex4-1}
\usepackage[dvips]{graphics,graphicx}
\usepackage{epstopdf}
\usepackage{bbold}
\usepackage[usenames, dvipsnames]{color}
\usepackage{lipsum}
\usepackage{amssymb}

\begin{document}
\title{Bound States in the Continuum and Fano Resonances in the Dirac Cone Spectrum}
\author{Evgeny N. Bulgakov$^{1,2}$}
\author{Dmitrii N. Maksimov$^{1,2,3}$}
\affiliation{$^1$Reshetnev Siberian State University of Science and Technology, 660037, Krasnoyarsk,
Russia\\
$^2$Kirensky Institute of Physics, Federal Research Center KSC SB
RAS, 660036, Krasnoyarsk, Russia\\
$^3$3
Siberian Federal University, Krasnoyarsk, 660041, Russia}
\date{\today}
\begin{abstract}
We consider light scattering by two dimensional arrays of high-index dielectric spheres arranged into the triangular
 lattice. It is demonstrated that in the case a triple degeneracy of resonant leaky modes in the Gamma-point the
 scattering spectra exhibit a complicated picture of Fano resonances with extremely narrow line-width. The Fano
 features are explained through coupled mode theory for a Dirac cone spectrum as a signature of optical bound states in the continuum (BIC).
 It is found that the standing wave in-Gamma BIC induces a ring of off-Gamma BICs due to different scaling laws for real and
 imaginary parts of the resonant eigenfrequencies in the Dirac cone spectrum.
  A quantitative theory
 of the spectra is proposed.
\end{abstract}
 \maketitle


\section{Introduction}

Photonic band engineering plays important role in modern science and technology \cite{Ding04, Joannopoulos11}.
Among numerous implementations the attention has been payed to photonic crystalline designs supporting
Dirac cone spectrum about the $\Gamma$-point which pave a way to all-dielectric
zero-refractive-index materials \cite{Huang11, Chan12}. Recently zero-index all-dielectric metamaterials have
been proposed \cite{Minkov18} relaying on the effect of optical bound states in the continuum (BICs) which are
lossless localized solutions coexisting with the continuous spectrum of the scattering states \cite{Hsu16}.
The emergence of BICs is remarkable due to their effect on the scattering of electromagnetic waves. The BICs
are known to induce sharp Fano \cite{Lee12,Kim,Shipman,SBR,Blanchard16} resonances in the scattering spectra due to interference between two optical
pathways, the resonant pathway via the subradiant mode associated to the BIC and the direct pathways due to the non-resonant scattering
\cite{Bulgakov2015, Blanchard16, Bulgakov18b, Bochkova18, Bogdanov19}. In principle in the spectral vicinity of a BIC the Q-factor of the
the resonances can be tuned to arbitrary high value once the material losses are neglected \cite{Yuan17,Bulgakov18c}.
In this paper we examine the effect of the radiation losses on the spectrum of the leaky bands and Fano resonances about the Dirac
point in a dielectric structure extended in two dimensions.

\section{Spectrum of leaky modes and Fano resonances}

\begin{figure}[t]
\centering
\fbox{\includegraphics[width=0.4\textwidth,height=0.35\textwidth,trim={7cm
11.5cm 6cm 11.cm},clip]{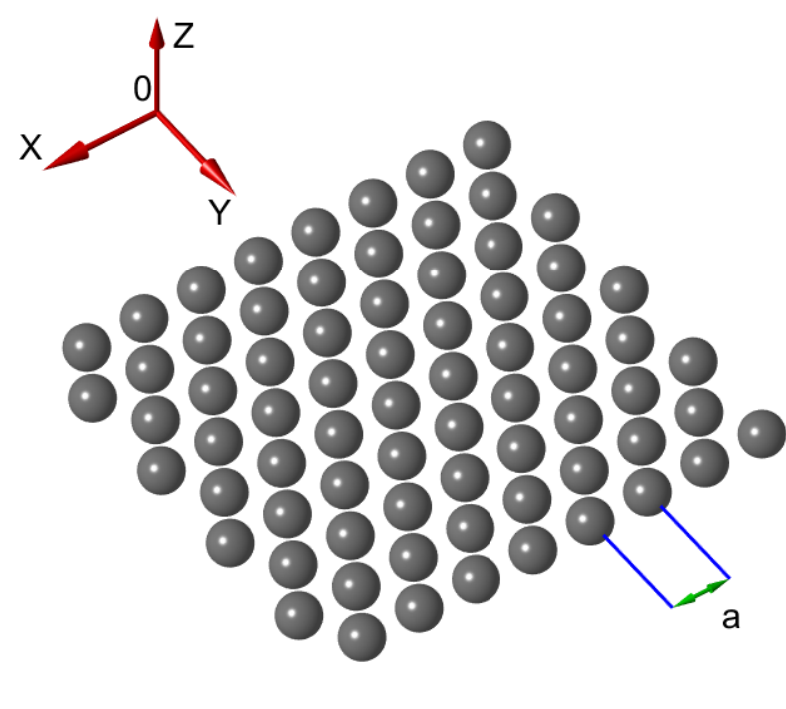}} \caption{Triangular lattice of
dielectric spheres.} \label{fig1}
\end{figure}

We consider a periodic two-dimensional array of high-contrast
($\epsilon=15$) dielectric spheres of radius $R$ arranged into a triangular
lattice in $x0y$-plane with period $a$ as shown in Fig. \ref{fig1}. Further on the frequency
will be expressed in terms of the vacuum wavevector $k_0$. To recover the
optical properties of the system we employ The
Korringa-Kohn-Rostoker method that was adapted  to scattering of
EM waves by 2D arrays of dielectric spheres by Ohtaka
\cite{Ohtaka79, Ohtaka80}. The method was later generalized for
finding the band structure \cite{Ohtaka00}.

The triangular lattice complies with $C_{6v}$ point symmetry group
\cite{Inui12} which has four one-dimensional $(A_1, A_2, B_1,
B_2)$ and two two-dimensional representations $(E_1, E_2)$.
\begin{figure*}[t]
\centering
\fbox{\includegraphics[width=0.9\textwidth,height=0.6\textwidth,trim={0.5cm
8.0cm 2cm 8.5cm},clip]{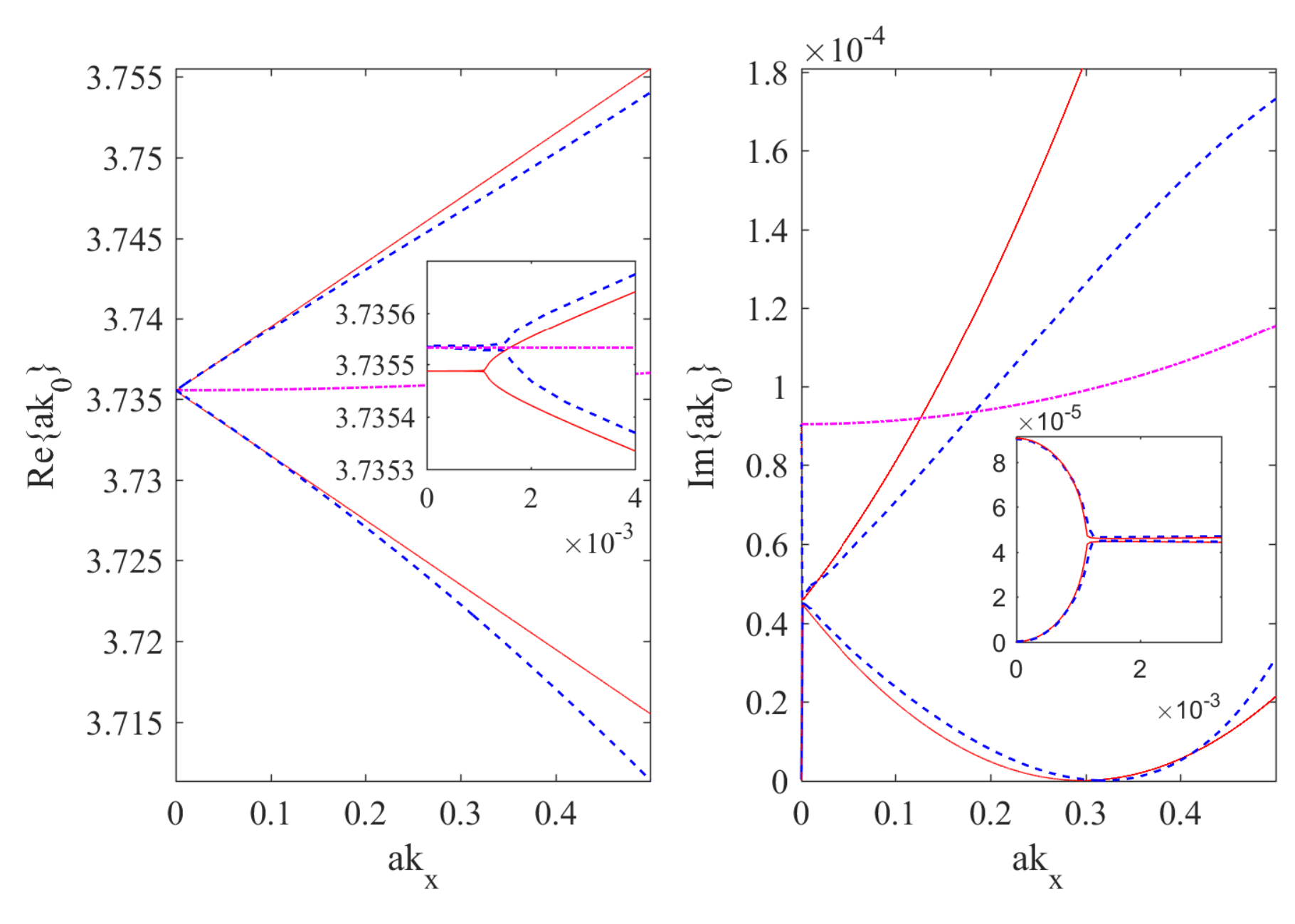}} \caption{The spectrum of leaky bands in the vicinity of the $\Gamma$-point. The real part of the
eigenfrequency - left panel, the imaginary part - right panel. The insets resolve the spectra to $ak_x\approx 10^{-3}$.
The dash blue lines show the spectrum of the hybridized modes, dash-dot magenta - $p_y$-mode. The red lines are eigenvulues
of the matrix Eq. (\ref{Heff}). } \label{fig2}
\end{figure*}
According to Sakoda \cite{Sakoda12}  a Dirac cone can be
engineered through a triple degeneracy between two modes, $p_x$
and $p_y$, of irreducible representation $E_1$,  and one mode,
$s$, of irreducible representation $A_1$ in the $\Gamma$-point.
$p_x$ and $p_y$ have the same freguency $\Omega_1$ in the
$\Gamma$-point, since they are of the same irreducible
representation. The $s$ mode, though, generally has a different
frequency $\Omega_2$. The degeneracy $\Omega_1=\Omega_2$ is
accidental in nature and be obtained by tuning the radius of the
spheres. In case $\epsilon=15$ the degeneracy is found at
$R=0.4705a$. Once the degeneracy is achieved the spectrum in the
vicinity of the $\Gamma$-point consists of an isotropic Dirac cone
and a quadratic dispersion surface \cite{Sakoda12}.

The spectrum of the leaky modes against the $x$-component of the wavevector, $k_x$ is
shown in Fig. \ref{fig2}. One can seen in Fig. \ref{fig2} that the real parts of the eigenfrequencies
demonstrate a behavior very close to that predicted in \cite{Sakoda12} with two bands forming
a Dirac cone while the third, weakly dispersive band is parabolic. The Dirac spectrum in $x0z$ plane
is formed by hybridization between $s$- and $p_x$-modes which both have their
electric fields antisymmetric with respect to $y\rightarrow -y$. Notice that
the radiation losses result in an anomaly in the spectrum with the dispersion deviating from a Dirac cone at small values of
$k_x$ as it is seen from the insets. That anomaly associated with an exceptional point
has been considered in detail by Zhen and co-authors \cite{Zhen15a}.

More interesting, though, is the behavior of the imaginary parts of the eigenfrequency which correspond
to the widths of the resonances. In the $\Gamma$-point the low frequency band obtains zero
imaginary parts which indicates the emergence of symmetry protected BIC - standing wave decoupled from
outgoing waves by symmetry \cite{Minkov18}. The other two modes, $p_x$ and $p_y$ are leaky in the $\Gamma$-point. With a slightest
off-set in the $k$-space the imaginary pars undergo dramatic changes due to hybridization between $s$- and $p_x$-modes.
That results in almost equal imaginary parts at $ak_x\approx 10^{-3}$. Then the imaginary part of the high frequency band
gradually increased with $k_x$, while the imaginary part of of the low frequency band drops to zero at $ak_x\approx 0.317$
where we find the second BIC. Since the BICs of that type have non-zero wave vector along the axis of periodicity of
the structure, they are termed Bloch BICs \cite{Bulgakov2015}.

\begin{figure*}[t]
\centering
\fbox{\includegraphics[width=0.9\textwidth,height=0.47\textwidth,trim={0cm
8.5cm 0cm 10cm},clip]{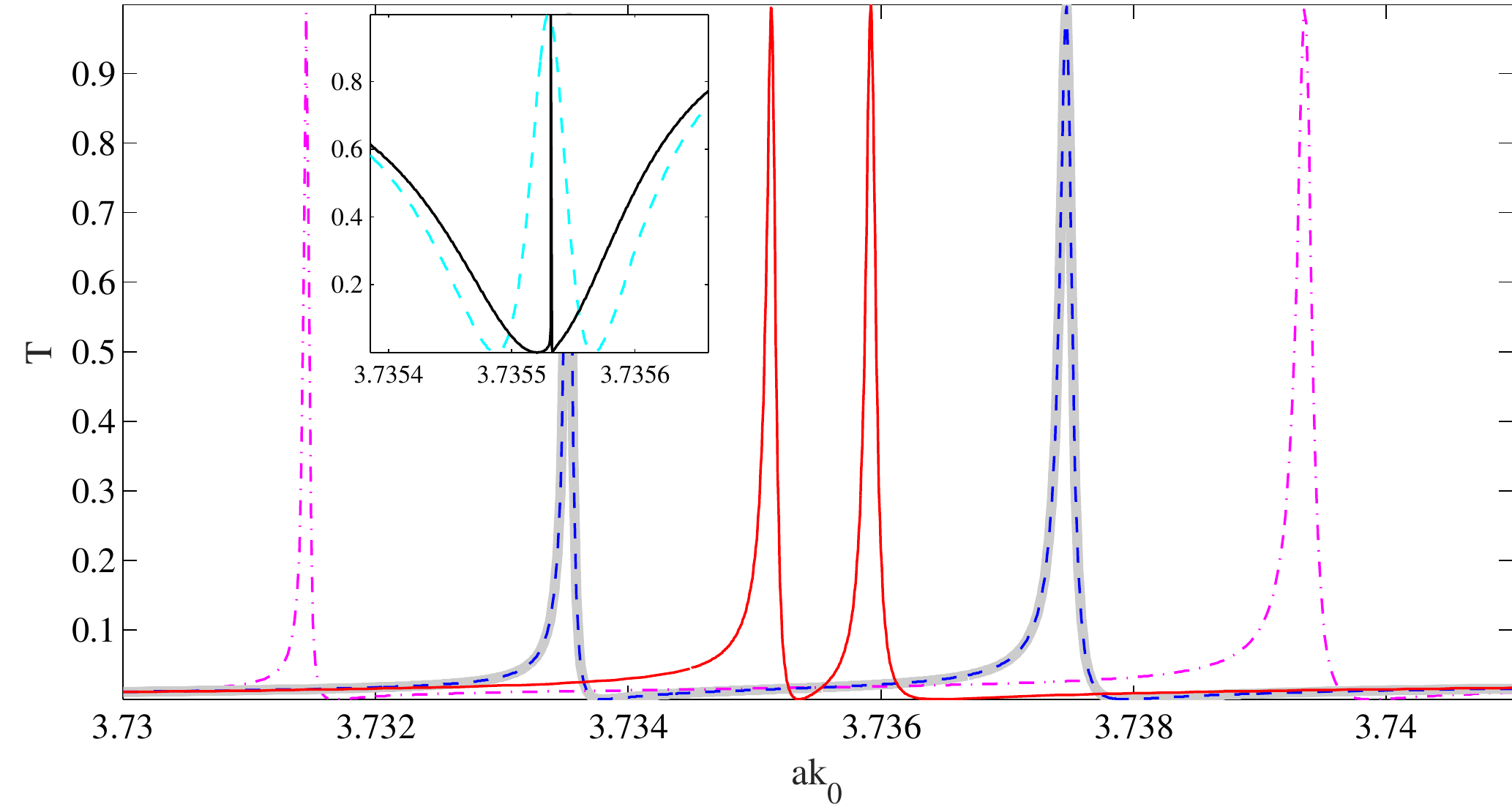}} \caption{Transmittance spectrum of the $TE$-wave, $k_y=0$. Red line - $ak_x=0.01$, blue dash line
- $ak_x=0.05$, dash-dot magenta - $ak_x=0.1$. The inset shows the resonant feature in the vicinity of the exceptional point,
teal dash line - $ak_x=10^{-3}$, solid black - $ak_x=8\cdot10^{-5}$. The thick gray line shown the case of the $TE$-wave with $ak_x=0, ak_y=0.05$. } \label{fig3}
\end{figure*}
Next we propose a simple phenomenological approach explaining the structure of the spectrum
in Fig. \ref{fig2}. According to \cite{Sakoda12} the spectrum of hybridized modes of $E_1$ and $A_1$ representations is found as the
eigenvalues of the "Hamiltonian" matrix
\begin{equation}
C_0= \left(
\begin{array}{ccc}
0 & 0 & bk_x \\
0 & 0 & bk_y \\
b^{*}k_x & b^{*}k_y &0
\end{array}
\right), \label{bigmatrix}
\end{equation}
where $b$ is a constant which can be evaluated by solving
Maxwell's equations numerically \cite{Sakoda12}. Assume that the
propagation direction of the incident wave is orthogonal to the
$y$-axis, i.e. $k_y=0$. Then Eq. (\ref{bigmatrix}) is reduced to
$2\times 2$ matrix
\begin{equation}
H_0= \left(
\begin{array}{cc}
0 & bk_x \\
b^{*}k_x & 0
\end{array}
\right), \label{smallmatrix}
\end{equation}
The radiation losses can be incorporated to Eq.
(\ref{smallmatrix}) by applying coupled mode theory \cite{Suh04}
for the hybridized resonances in the following manner
\begin{equation}
H=H_0+\frac{i}{2}W^{\dagger}W, \label{Heff}
\end{equation}
where $W^{\dagger}=(\sqrt{\gamma_p}, \sqrt{\gamma_s})$ with $\gamma_s$ and $\gamma_p$ being
the decay rates of the hybridizing $s$- and $p_x$-modes into the $TE$-wave radiation channel with $TE$-outgoing (incoming) wave being
defined as that whose electric vector is perpendicular to the plane of incidence.  Notice
that for $k_y=0$ the hybridized modes are decoupled from the $TM$-wave by symmetry.  Having in mind the
definition Eq. (\ref{Heff}), the quantity
$\gamma_p$ can be found from Fig. \ref{fig2} as one half of the imaginary part of the resonant eigenfrequency of the $p_x$-mode in the
$\Gamma$-point. On the other hand, $\gamma_s$ in the vicinity of the $\Gamma$-point must be a
dispersive quantity to reflect the singular nature of the symmetry protected BIC \cite{Bulgakov18b}.
The leading term in the $k_x$ expansion of $\gamma_s$ is quadratic since $k_x=0$ is absolute minimum
of the line width corresponding to the symmetry protected BICs. The dispersion of $\gamma_s$ can be accessed
by slightly detuning the radius $R$ to lift the degeneracy so the individual features of the $s$-mode can be resolved.
By running simulations with $R=0.4514$ one finds $a\gamma_s=1.028\cdot 10^{-3}\cdot k_x^2$. Finally, the parameter $b$ can be found
via perturbative approach in \cite{Sakoda12} or simply extracted from Fig. \ref{fig2} as the real parts repulsion rate away off
the feature at $ak_x\approx 10^{-3}$. Here we found $b=0.04$.
In Fig. \ref{fig2} we demonstrate the spectrum of the matrix Eq. (\ref{Heff})  that is found to be in quantitative agreement
with numerical data. One can see that the model predicts the emergence of the satellite Bloch BIC as well as the anomalous spectral
feature in the immediate vicinity of the $\Gamma$-point.

Now, let us discuss the effect of the spectra on the scattering. It is known that the presence of high-Q leaky modes results
in narrow Fano feature which collapse as the spectral parameters are tuned to a BIC \cite{Bulgakov2015,Blanchard16, Bulgakov18b, Bogdanov19}.
In our situation away off the exceptional point the $TE$-wave scattering should reveal two isolated Fano resonances whose positions are given by the real parts of the
Dirac cone eigenfrequencies while the width is controlled by the imaginary part of the spectrum. That statement is in full agreement
with the numerical data shown in Fig. (\ref{fig3}). In the vicinity of the exceptional point the Fano resonances merge into a single feature
which can be only resolved on zoomed scale in $k_0$ as shown in the inset in Fig. (\ref{fig3}). Finally, we mention in passing that
$TM$-waves are only coupled to the single parabolic band resulting to a single, almost non-dispersive Fano feature.

\section{Ring of BICs}

What is remarkable according to Eq. (\ref{bigmatrix})
the Dirac cone is isotropic in the momentum space in the vicinity of the $\Gamma$-point. Let us see whether the presence of the radiation
losses breaks that symmetry. Let specify the propagation direction specified by arbitrarily azimuthal angle $\phi$ such that
$k_x=k_{\parallel}cos(\phi)$ and $k_y=k_{\parallel}sin(\phi)$ with  $k_{\parallel}=\sqrt{k_x^2+k_y^2}$.
The Hamiltonian has to be written in the following form  \cite{Suh04}
\begin{equation}
C=C_0+\frac{i}{2}W^{\dagger}W
\label{Heff2}
\end{equation}
where $W$ is a $2\times 3$ matrix composed of the row matrices, $W=[W_{TE}, W_{TM}]$, that describe the coupling with $TE$ and $TM$-waves, correspondingly.
Taking into account that $p_x \rightarrow p_y$
and $p_y\rightarrow -p_x$ under
rotation by $\phi=\pi$ we can write
\begin{eqnarray}
W_{TE}=[\sqrt{\gamma_p}\cos(\phi), \sqrt{\gamma_p}\sin(\phi), \sqrt{\gamma_s}], \\
W_{TM}=[-\sqrt{\gamma_p}\sin(\phi), \sqrt{\gamma_p}\cos(\phi), 0].
\end{eqnarray}
Next, one can easily check that after transforming the Hamiltonian $C'=R^{\dagger}CR$ with matrix $R$
\begin{equation}
R= \left(
\begin{array}{ccc}
\cos(\phi) & -\sin(\phi) & 0 \\
\sin(\phi) & \cos(\phi) & 0 \\
0 & 0 &1
\end{array}
\right), \label{transformation}
\end{equation}
one finds
\begin{equation}
C'= \left(
\begin{array}{ccc}
0 & 0 & bk_{\parallel} \\
0 & 0 & 0 \\
b^{*}k_{\parallel} & 0 & 0
\end{array}
\right)+\frac{i}{2}
\left(
\begin{array}{ccc}
\gamma_p & 0 & \sqrt{\gamma_s \gamma_p}\\
0 & \gamma_p & 0 \\
 \sqrt{\gamma_s \gamma_p} & 0 &\gamma_s
\end{array}
\right),
\label{bigmatrix_trans}
\end{equation}
which reduces to Eq. (\ref{Heff}) in terms of coupling to the $TE$-wave. Moreover,
one can immediately see that the spectrum of complex eigenfrequencies is invariant under rotation.
It means that our observation on the $TE$-wave scattering
apply for any orientation of the plane of incidence given by $\phi$. This
statement is illustrated in Fig. (\ref{fig3}) for the incident wave with $k_x=0, ak_y=0.05$.
Since the two-sate model applies for any $\phi$ the frequencies of the satellite Bloch BIC form
a ring around the $\Gamma$-point.

\begin{figure}[t]
\centering
\fbox{\includegraphics[width=0.43\textwidth,height=0.23\textwidth,trim={0cm
8.5cm 1cm 11.5cm},clip]{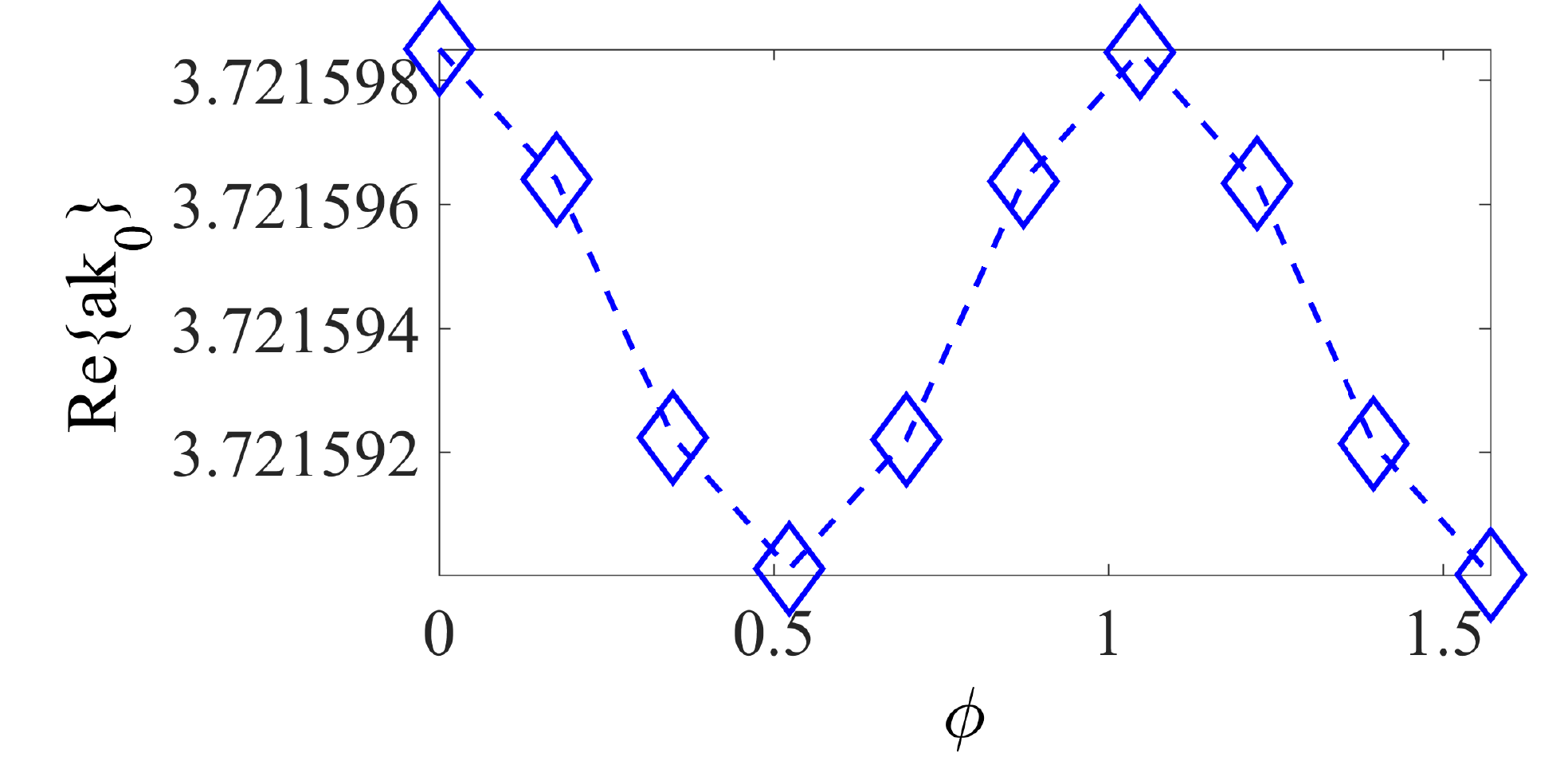}} \caption{Eigenfrequencies of the off-$\Gamma$ BICs as the function of the azimuthal angle $\phi$. } \label{fig4}
\end{figure}

\begin{figure}[t]
\centering
\fbox{\includegraphics[width=0.43\textwidth,height=0.76\textwidth,trim={7cm
8cm 7cm 8cm},clip]{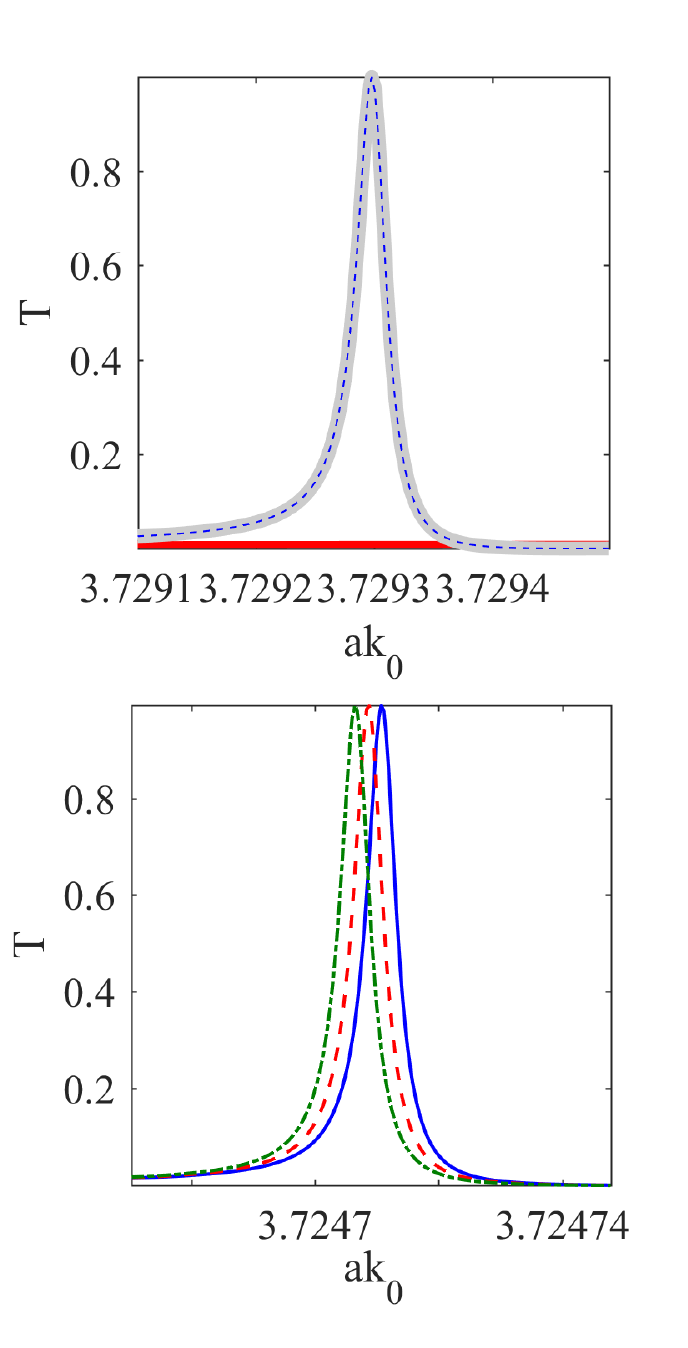}} \caption{Transmittance spectrum of $TE$-waves in the vicinity of the ring of BICs. Upper panel, $ak_{\parallel}=0.15$; thick
gray line - $\phi=0$, thin dash blue - $\phi=\pi/2$. Solid red line is the transmittance of $TM$-waves.
Lower panel, $ak_{\parallel}=0.25$; solid blue
- $\phi=0$, dash red - $\phi=\pi/4$, dash-dot green - $\phi=\pi/2$. } \label{fig5}
\end{figure}

In Fig. \ref{fig4} we show the eigenfrequencies of the of the off-$\Gamma$ BICs as a function of the azimuthal angle $\phi$.
One can see from Fig. \ref{fig4} that the BICs form almost ideally circular ring with the lattice anisotropy footprint emerging only
in the seventh significant digit. Let us consider the transmittance spectrum in the immediate vicinity of the off-$\Gamma$ BICs.
Assume that the frequency of the incident wave is tuned to  the lowest branch in Fig. \ref{fig2}. Then, if $k_{\parallel}$ of the incident
is matched to that of the leaky mode, the scattering spectrum of $TE$-waves exhibits an extremely narrow Fano feature whose profile is almost
independent on the azimuthal angle. This is illustrated in Fig. \ref{fig5} (upper panel) for $ak_{\parallel}=0.15$. In the same
subplot we also demonstrate that the resonant feature can be only observed with $TE$-waves while the transmittance of the $TM$-waves
remains independent of frequency on the scale of the Fano resonance. Finally, on the further approach to the
ring of BICs the lattice anisotropy emerges as a small shift of the resonance positions with respect to each other as seen from
Fig. \ref{fig5} (lower panel).

\section{Conclusion}

In summary, here we went beyond the ring of exceptional points predicted in \cite{Zhen15a} by taking into account
the dispersion of the hybridized modes line widths. We have demonstrated that the presence of a Dirac cone in the $\Gamma$-point results in the emergence
of high-Q Fano resonances in the transmittance spectrum of $TE$-waves near the normal incidence. The positions
and widths of the resonances can estimated from a simple coupled mode approach leading to $2\times2$ matrix whose
parameters are easily extracted form the dispersion of the hybridized leaky modes. It is shown that in the vicinity of
the $\Gamma$-point the scattering spectra are insensitive to the orientation of the plane of incidence. Most remarkably,
it is found that the presence of a Dirac cone together with an in-$\Gamma$ symmetry protected BIC induces a ring
of Bloch BICs surrounding the $\Gamma$-point the $k$-space. The emergence of the BICs in the model is a result
of destructive interference between two resonant modes with the condition for BICs \cite{Volya03} being fulfilled by
tuning the wavenumber due to the difference in asymptotic behavior between Hermitian and non-Hermitian parts of the matrix
in Eq. (\ref{Heff}).
The presence of such BICs allows for fine tuning the width
of Fano resonances by changing the angle of incidence. Recently, we have seen a big deal of interest to application of Fano
resonances to sensing and switching \cite{Heuck13, Limonov17, Zhang18}. We believe that the proposed model
may be a useful platform for engineering Fano resonances in all-dielectric set-ups \cite{KRASNOK17}.
The possibility of BICs eigenfrequencies forming a ring in momentum space
has been recently pointed out in a view of multipolar decompositions \cite{Sadrieva19}. We speculate that a further detailed analysis
of multipolar structure could shed light onto interference mechanisms leading to BICs.

{\bf Funding.} This work was supported by Ministry of Education
and Science of Russian Federation (state contract N
3.1845.2017/4.6).

\bibliography{BSC_light_trapping}


\end{document}